\documentclass[9pt,twocolumn,twoside]{opticajnl}
\journal{opticajournal} 

\setboolean{shortarticle}{false}

\usepackage{lineno}
\usepackage{subfigure}

\newcommand{\im}{\mathrm{i}} 
\newcommand{\diff}{\mathrm{d}} 

\title{Optical enhancement cavities operated at GHz repetition rate in burst mode regime}

\author[1,*]{Kevin Dupraz}
\author[1,2]{Frédéric Blanc}%
\author[1]{Ronic Chiche}%
\author[1]{Aurélien Martens}%
\author[2]{Antoine Courjaud}
\author[1]{Lo\"ic Amoudry}%
\author[1]{Titouan Le Barillec}%
\author[1]{Daniele Nutarelli}%
\author[1]{Yann Peinaud}%
\author[1]{Fabian Zomer}%
\affil[1]{Université Paris-Saclay, CNRS/IN2P3, IJCLab, 91405 Orsay, France.}
\affil[2]{Amplitude, 11, avenue de Canteranne, Cité de la Photonique Bâtiment MEROPA, 33600 Pessac, France}

\affil[*]{kevin.dupraz@ijclab.in2p3.fr} 

\begin{abstract}
The design and experimental validation of a burst-mode optical cavity tailored for inverse Compton scattering X-ray sources is presented. The system is tailored for a room temperature linear accelerator operating at 880~MHz and utilizes a common-path injection scheme for both locking and burst laser pulses to optimize stability and simplicity. We detail the cavity construction, the laser system, and performance under two enhancement cavity finesse regimes (2000 and 7300), highlighting the sensitivity of cavity filling to carrier-envelope offset parameters. A theoretical model is developed to explain observed power dynamics and to guide optimization for high-energy storage. Our findings demonstrate reliable storage exceeding 300~mJ over 20~minutes, paving the way toward compact, high-flux inverse Compton scattering X-ray sources.
\end{abstract}

\setboolean{displaycopyright}{false} 

\begin{document}
\maketitle

\section{Introduction}
Inverse Compton X-ray sources (ICS)~\cite{ICS_initial,ICS_initial2} are designed to bridge the gap between large-scale synchrotron facilities, which provide the highest-brightness and most monochromatic X-ray beams, and laboratory-scale X-ray tubes, which suffer from limited flux, monochromaticity, and coherence~\cite{JACQUET_2016}. While numerous ICS facilities have been developed worldwide, only a limited number are dedicated to user-oriented applications requiring stable operation and high X-ray flux.

ICS architectures can be classified according to both the electron accelerator technology and the optical system employed at the interaction point. The highest X-ray fluxes are typically achieved using electron storage rings combined with high-finesse Fabry-Perot cavities~\cite{Huang_Ruth}, which provides both high laser power and high repetition rates~\cite{MuCLS_2020,ThomX_2025,Hajima_2008}. However, this configuration is not optimal in terms of compactness, especially for producing high-energy X-rays.

To reduce the facility footprint while maintaining high beam quality, linac-based ICS designs are particularly attractive. Fabry-Perot cavities coupled with superconducting linacs~\cite{BriXS_2020} provide maximum energy recycling and can achieve very high average photon flux. Nevertheless, superconducting and cryogenic technologies generally require substantial infrastructure, which limits overall compactness.

Modern room temperature linacs offer an alternative approach. Typical operating parameters involve radiofrequency pulses at repetition rates ranging from 100~Hz to 1~kHz, with pulse durations of a few microseconds~\cite{STAR_2016}. Since the accelerating structures operate at radiofrequencies in the gigahertz range, electron bunches can be generated at similar repetition rates, allowing thousands to tens of thousands of bunches to be accelerated within a single burst~\cite{Radiabeam_gun,SmartLight_gun}.

The optical system must therefore be adapted to this burst-mode operation. Several solutions have been investigated, including single-shot laser systems~\cite{TTX_2013,STAR_2016,Lumitron_2024,SmartLight_2025}, multipass and optical gating schemes~\cite{Ebina_2005,Graves_2014,Jovanovic_2007,ELI_2019}, and burst-mode enhancement cavities~\cite{Sakaue_2009,Sakaue_2018,Rakhman_burst_application}. Single-shot systems provide the simplest implementation but offer no energy recycling, whereas multipass and gated schemes enable partial reuse of the laser energy. Enhancement cavities provide the highest recycling efficiency, in such systems, high X-ray flux, compactness, and very high spectral purity can be obtained~\cite{Martens_2021}.

Despite these advantages, only a limited number of studies have investigated enhancement cavities specifically designed for burst-mode operation~\cite{Sakaue_2009}. Existing demonstrations have shown promising results~\cite{Sakaue_2018,Rakhman_burst_application}, but none were optimized for the combination of gigahertz bunch repetition frequencies and microsecond-long bursts required by modern room temperature linacs. For instance, Ref.~\cite{Sakaue_2018} reports a burst duration of 300~$\mu$s at a pulse repetition frequency of 357~MHz, while Ref.~\cite{Rakhman_burst_application} presents a two-wavelength enhancement cavity operated with burst durations ranging from tens of microseconds to milliseconds and a pulse repetition frequency of 402.5~MHz. In comparison, the cavity presented in this work is specifically optimized for a burst duration of 5~$\mu$s and a pulse repetition frequency of 880~MHz. Such an optimization was proposed theoretically in Ref.~\cite{Favier_2018}.

In this paper, we report the first experimental demonstration of an optical cavity specifically designed for a burst-mode linac operating at gigahertz repetition rates with burst durations on the order of a few microseconds. First, Section~\ref{sec:experimental_setup} describes the experimental setup together with the injection and locking schemes used to operate the cavity. Section~\ref{sec:Results_high_power} then presents experimental results obtained with a low-finesse cavity, demonstrating stable long-term storage of high burst energy. Finally, Section~\ref{sec:CEP_experiment} investigates the role of carrier-envelope offset (CEO) related parameters using a higher-finesse cavity. A simple model describing the cavity filling dynamics in the presence of CEO effects is introduced and compared with the experimental observations.

\section{Experimental setup and related considerations}\label{sec:experimental_setup}
    This section describes the experimental setup used for the burst storage experiments, including the injection and locking schemes employed for the two optical cavities implemented along the course of this work. Particular attention is given to the implementation of the common-path locking architecture. The section concludes with a discussion of the main experimental considerations and design constraints that must be taken into account when developing a burst-mode enhancement cavity.
    \subsection{Experimental setup and laser burst injection scheme}
\begin{figure}[htbp!]
            \centering
            \includegraphics[width=\linewidth]{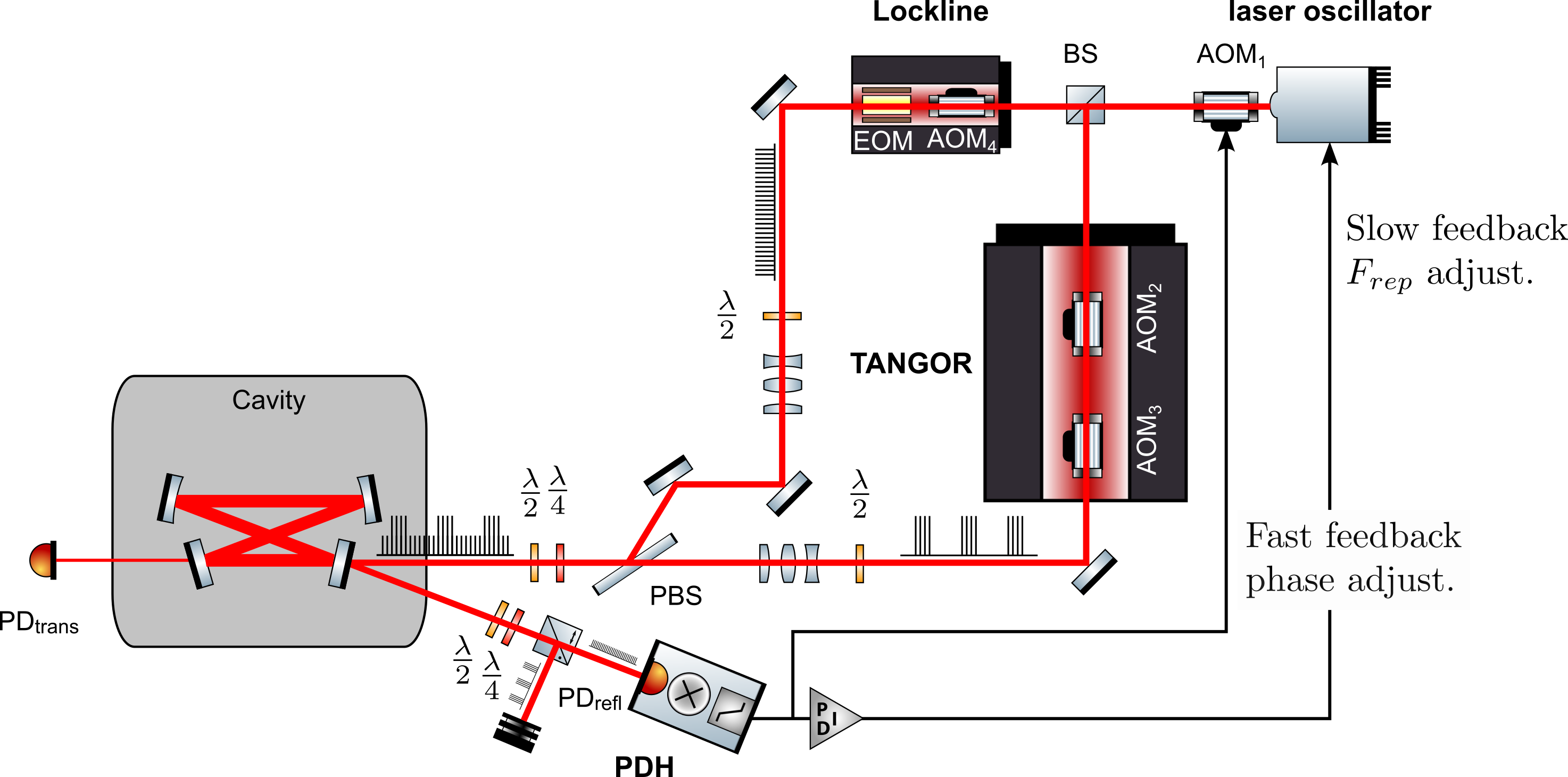}
            \caption{Schematic of the experimental setup. PBS: polarizing beam splitter; BS: beam splitter; $\text{AOM}_\text{i}$: acousto-optic modulator numbered i; EOM: electro-optic modulator; PD: photodiode; PDH: Pound-Drever-Hall detector; $\lambda / 2$: half wave plates; $\lambda / 4$: quarter wave plates, see text for details.}
            \label{fig:setup_minicav}
        \end{figure}

        The experimental setup used for the burst storage experiments presented in this work is largely based on the configurations reported in Ref.~\cite{Sakaue_2018, CW_burst}, with one key difference: we did not employ a counter-propagating beam for cavity locking. Instead, we adopted a common injection path for both the pulsed laser beam continuously injected for feedback (referred to as the Lockline) and the high-energy burst laser beam (the TANGOR line), see Fig.~\ref{fig:setup_minicav}. This technical choice was primarily motivated by the following considerations:

        \begin{itemize}
          \item The ability to isolate the optical reflections from each line (contrast ratio), thereby avoiding potential damage to the laser amplifiers caused by back-reflections. Achieving this in a dual-path system would have required two high-contrast optical isolators, with at least 53~dB of isolation for a 100~mW Lockline and a 100~W TANGOR burst line with a 5~$\mu$s duration at a 1 kHz burst repetition rate. The proposed scheme developed in this work avoids back-reflection at the input mirror. Thus, optical isolators are not intrinsically required.
          \item To avoid the complexity of maintaining independent alignment feedback loops for the two optical paths. This facilitates the transport of the laser beams over several meters for injecting in the optical cavity while the laser system usually lie on a table in the accelerator environment~\cite{MuCLS_2020,ThomX_2025,Sakaue_2018}.
          \item To avoid the need to duplicate all diagnostics: transmitted beam mode profiles, transmitted power, reflected power, etc.
        \end{itemize}

        A schematic of the optical setup is shown in Fig.~\ref{fig:setup_minicav}. Laser pulses are generated by a high-repetition-rate mode-locked oscillator operating at $f_{\text{rep}} = 880$~MHz. These pulses are evenly split into two separate paths. The first path feeds the TANGOR amplifier, a 100~W burst-mode power amplifier~\cite{TANGOR_2020}. The second path feeds the Lockline, a low-power amplifier (approximately 100~mW) operating in a steady pulse sequence. The complete laser system, including oscillator and amplifiers, is supplied by Amplitude Laser, using a combination of commercial and custom-built components.

        The train of pulses is shaped using two acousto-optic modulators (AOMs) to form a burst of laser pulses. The first: $\text{AOM}_\text{2}$, defines the amplification window on the continuous pulse train, while the second: $\text{AOM}_\text{3}$, selects specific amplified bursts and adjusts the output power. This second AOM is particularly useful for tuning the burst repetition rate and output power without altering the amplifier dynamics.

        The continuous pulse train (Lockline) is used to generate the error signal required to maintain the optical cavity at resonance. The locking technique is based on the Pound-Drever-Hall (PDH) method~\cite{Drever_PDH}. The necessary phase modulation of the optical wave is introduced via an electro-optic modulator (EOM) located within the Lockline amplifier chain. To compensate for the frequency shift introduced by the two AOMs ($\text{AOM}_\text{2}$ and $\text{AOM}_\text{3}$) in the TANGOR line, an additional AOM ($\text{AOM}_\text{4}$) is inserted into the Lockline amplifier.

        The two laser lines are then recombined using a polarizing beam splitter (PBS), with orthogonal polarizations. The polarization of the TANGOR beam is subsequently tuned to match the main cavity’s eigenmode polarization, i.e., the polarization of the Lockline beam is on the orthogonal mode.

        The PDH error signal is extracted from the reflected Lockline beam after passing through a polarization filter, which suppresses any residual contribution from the TANGOR beam. The signal is acquired using a photodiode (PD$_\text{refl}$) then demodulated. Once the resonance is crossed during a frequency scan, the transmitted power detected by another photodiode (PD$_\text{trans}$) triggers a proportional-integral-derivative (PID) feedback loop.

        It is worth noting that, even for a relatively low-finesse cavity, two feedback loops are required:
        \begin{itemize}
          \item A slow feedback loop to keep the laser repetition rate $f_{\text{rep}}$ synchronized with the cavity length.
          \item A fast feedback loop using an AOM ($\text{AOM}_\text{1}$) to compensate for oscillator phase noise.
        \end{itemize}

    \subsection{Optical cavity design}
        The design of the enhanced optical cavity was based on the optimization framework developed in Ref.~\cite{Favier_2018}. The following parameters were chosen as design constraints:

        \begin{itemize}
          \item A total of 1000 electron bunches per burst,
          \item A bunch repetition frequency $f_{\text{rep}}$ around 1~GHz; for this study, we used 880~MHz,
          \item Commercially available high-reflectivity mirrors,
          \item A symmetric bow-tie configuration composed of two planar and two spherical mirrors,
          \item A laser beam size of approximately 300~$\mu$m on the planar mirrors to reach easily the laser-induced damage threshold, for studying the mirrors limits.
        \end{itemize}

\begin{figure}[htb!]
                \centering
                \begin{minipage}{1\columnwidth}
                    \centering
                    \subfigure[]{
                    \includegraphics[width=\linewidth]{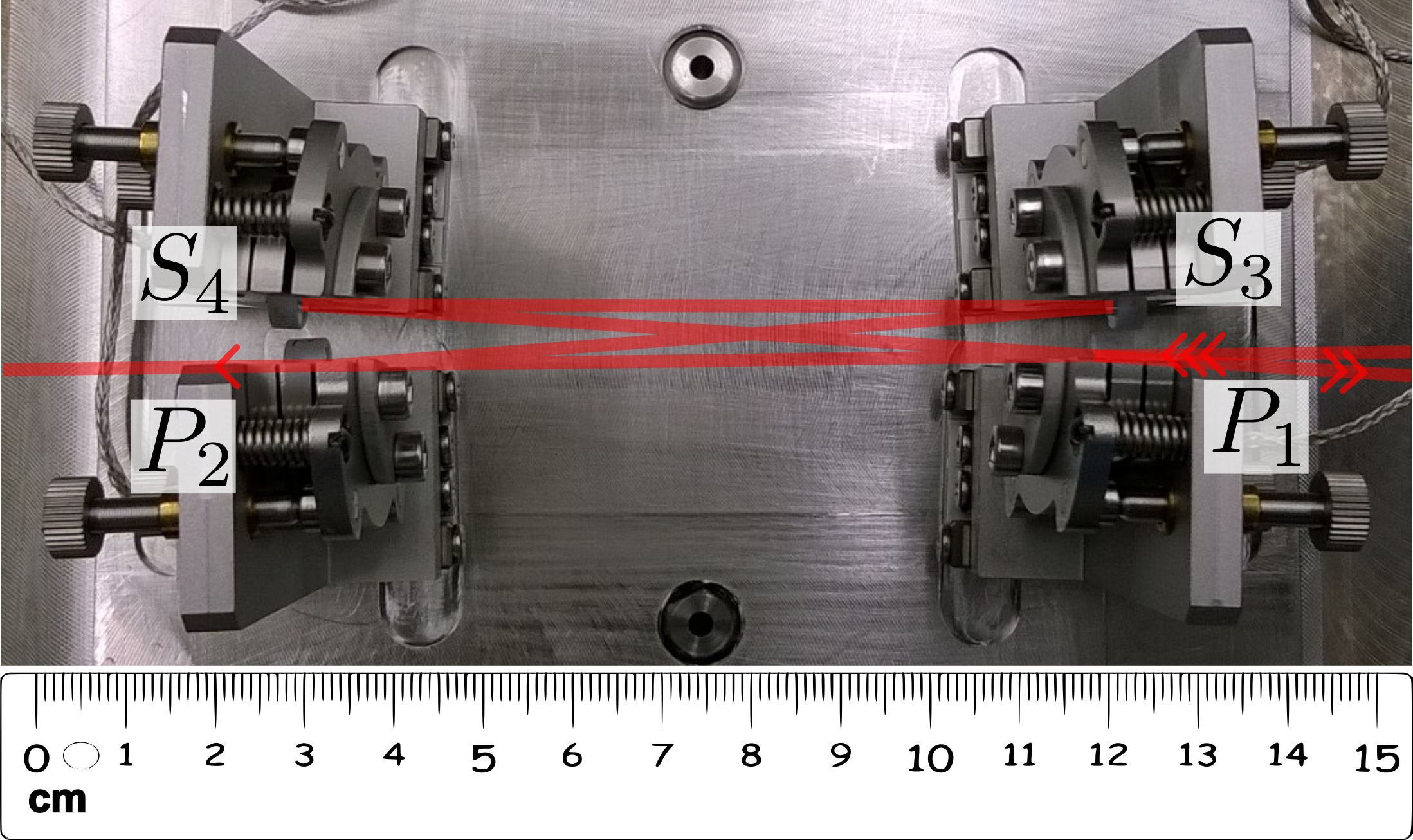}
                    \label{sfig:minicav_top_view}}
                \end{minipage}\\
                \begin{minipage}{1\columnwidth}
                    \centering
                    \subfigure[]{
                    \includegraphics[width=\linewidth]{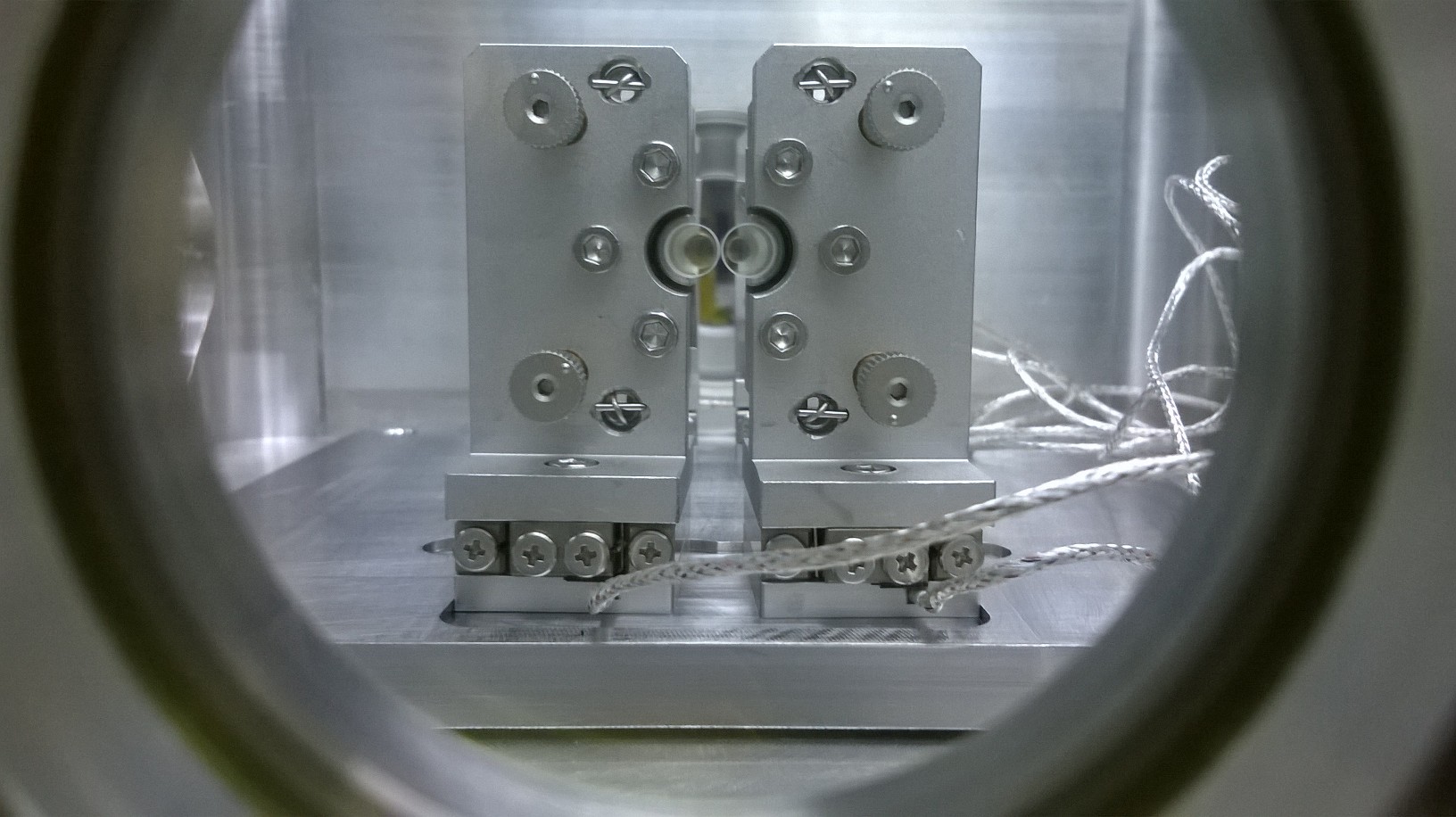}
                    \label{sfig:minicav_front_view}}
                \end{minipage}
                \caption{Photographs of the optical cavity. \subref{sfig:minicav_top_view} Top view with the optical path highlighted in red. The 1/4 inch mirrors are labeled $M_i$, where $M = S$ denotes spherical mirrors and $M = P$ denotes planar mirrors. The index $i$ corresponds to the mirror order along the propagation direction. \subref{sfig:minicav_front_view} Front view from the injection side. The distance between mirrors $S_4$ and $S_3$ is around to 90~mm.}
                \label{fig:minicav_pictures}
            \end{figure}

        These parameters resulted in the geometry illustrated in Fig.~\ref{fig:minicav_pictures}. The foreseen finesse is approximately 10,000 using standard commercial mirrors with a diameter of 1/4 inch. The optical path length between the two spherical mirrors is around 90~mm.

        As shown in Fig.~\ref{sfig:minicav_front_view}, the mirrors on one side of the cavity are positioned very close to each other. For operational flexibility, each mirror is mounted on a separate translation stage which allows independent tuning of the focal spot size and the cavity repetition frequency.

    \subsection{Experimental considerations} \label{ssec:Experimental_considerations}
        Several experimental considerations must be taken into account for the successful operation of the optical cavity.

        First, the target finesse could not be exactly achieved with commercially available mirrors, and therefore a design margin must be included. An uncertainty on the target finesse arises from the exact values of the mirror transmission, reflection, scattering, and absorption. The main contribution, however, comes from the final transmission of the coupling mirror, which is specified to achieve a given target finesse. As a result, the actual cavity measured finesse can differ from the target value, affecting key cavity parameters such as the cavity gain, the coupling (e.g., impedance matching and thermal sensitivity), and the sensitivity to the carrier-envelope offset (CEO) shift~\cite{Zomer_CEP}. This is why these are experimentally measured and not based on manufacturer's data. CEO shift effect becomes particularly important for high-finesse cavities, as discussed in Sec.~\ref{sec:CEP_experiment}.

        Second, attention must be paid to the extinction ratio of the polarization systems used to combine or separate the injected and reflected beams. We found that imperfections in the polarization optics could induce unwanted coupling between the TANGOR and Lockline beams. Such residual coupling cannot be filtered efficiently and may degrade the PDH error signal, potentially compromising cavity locking stability.

        Finally, the ratio between the mirror aperture and its thickness approaches unity in our setup, which leads to significant diffraction effects. As a result, beam path filtering is essential, and thermal management strategies must be implemented to handle the power deposited by the diffracted light.

\section{Results of high-power storage}\label{sec:Results_high_power}
    This section demonstrates the feasibility of efficiently storing high-energy laser bursts in an enhancement cavity operated in burst mode. Particular emphasis is placed on the long-term stability of the stored energy and on the achievable power enhancement, as these are key parameters for operation in a future accelerator-based facility. For this initial demonstration, a low-finesse cavity ($\mathcal{F}\approx2000$) was employed to benefit from a larger resonance linewidth and more robust operation. The injected burst energy was intentionally limited to remain well below the estimated laser-induced damage threshold of the cavity mirrors. Under these conditions, the stored energy scales approximately linearly with the injected energy, provided that thermal effects remain negligible. The section concludes with the main issues encountered when operating the cavity at higher finesse.
    \subsection{Power stored with a finesse of 2000}
        We successfully stored power within the cavity without major difficulty. The finesse of the test cavity was measured to be approximately $\mathcal{F} = 2000 \pm 200$, using the cavity sweeping method described in Ref.~\cite{Locke_09}. The transmission of the input coupling mirror $P_1$ (see Fig.~\ref{fig:minicav_pictures}) was measured as $T_1 = 590 \pm 30 \mathrm{ppm}$ (part per million), leading to an estimated power enhancement factor $G \approx \frac{T_1 \mathcal{F}^2}{\pi^2} = 240 \pm  50$. This gain was independently verified using the Lockline and a power meter placed behind mirror $P_2$. The transmission of the end mirrors ($P_2$, $P_3$ and $P_4$) was measured to be $11.9 \pm 0.4 \mathrm{ppm}$.

\begin{figure}[htb!]
                \centering
                \begin{minipage}{0.8\columnwidth}
                    \centering
                    \subfigure[]{
                    \includegraphics[width=\linewidth]{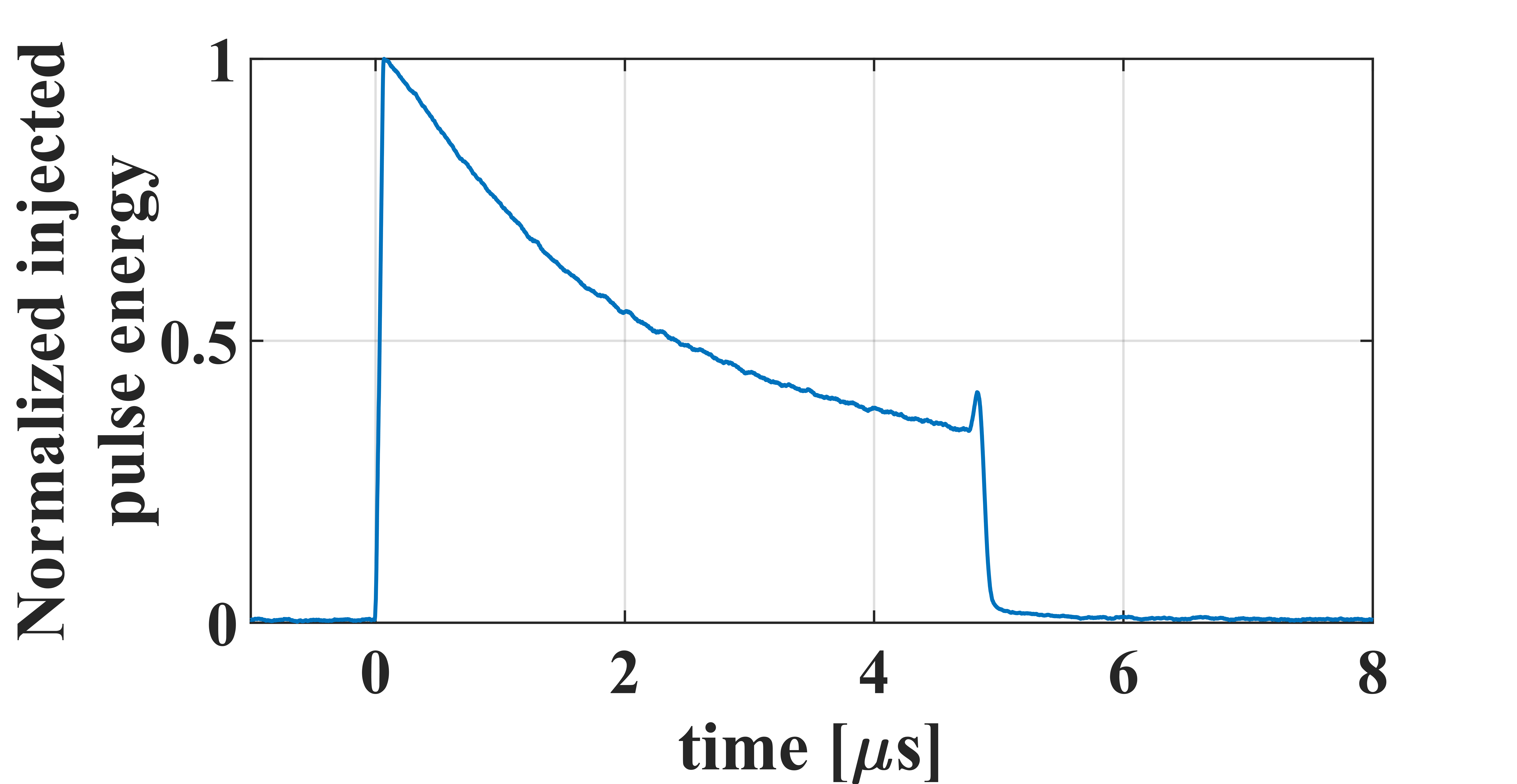}
                    \label{sfig:longterm_injected_burst}}
                \end{minipage}
                \begin{minipage}{0.8\columnwidth}
                    \centering
                    \subfigure[]{
                    \includegraphics[width=\linewidth]{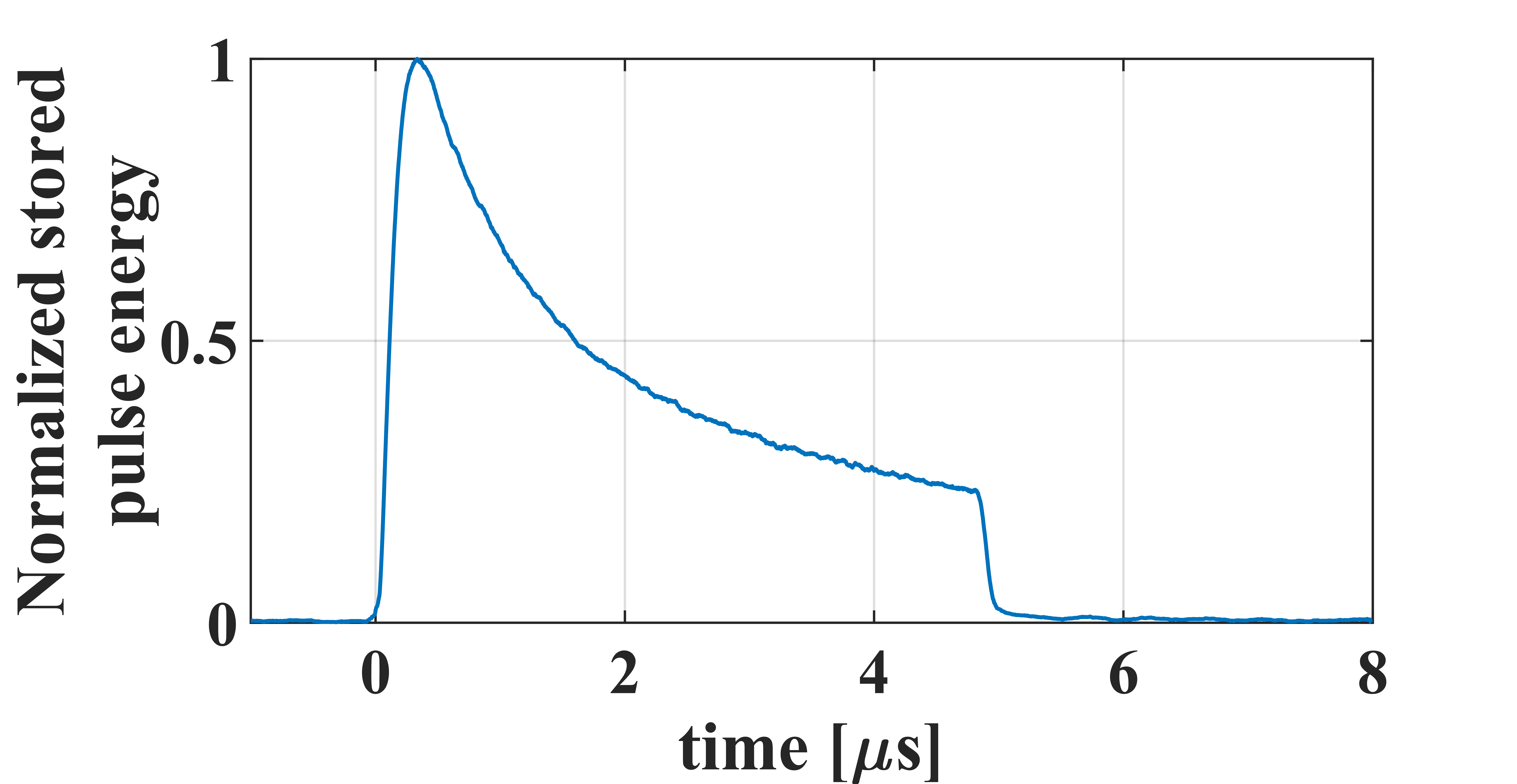}
                    \label{sfig:longterm_stored_burst}}
                \end{minipage}\\
                \begin{minipage}{0.8\columnwidth}
                    \centering
                    \subfigure[]{
                    \includegraphics[width=\linewidth]{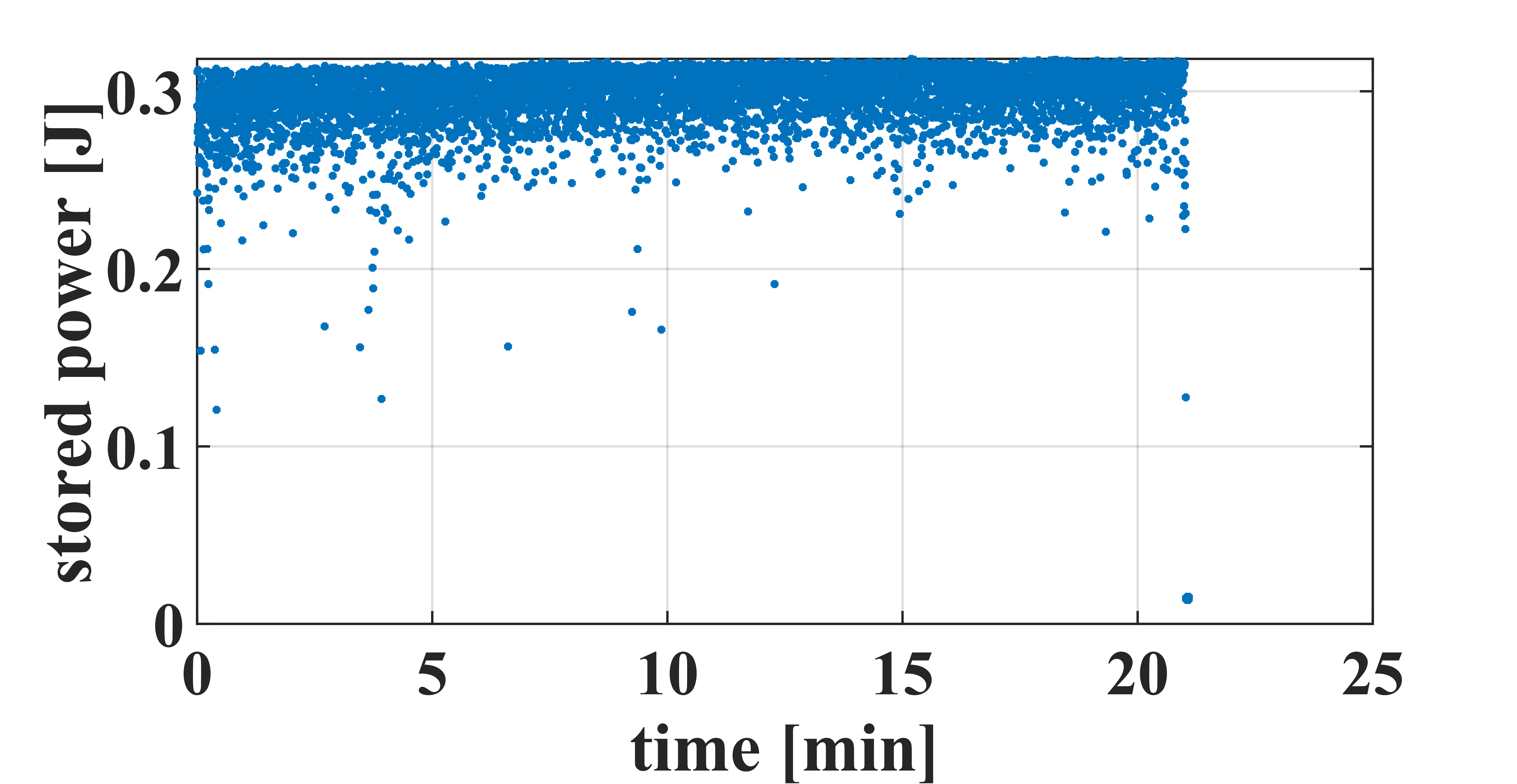}
                    \label{sfig:longterm_stored}}
                \end{minipage}
                \begin{minipage}{0.8\columnwidth}
                    \centering
                    \subfigure[]{
                    \includegraphics[width=\linewidth]{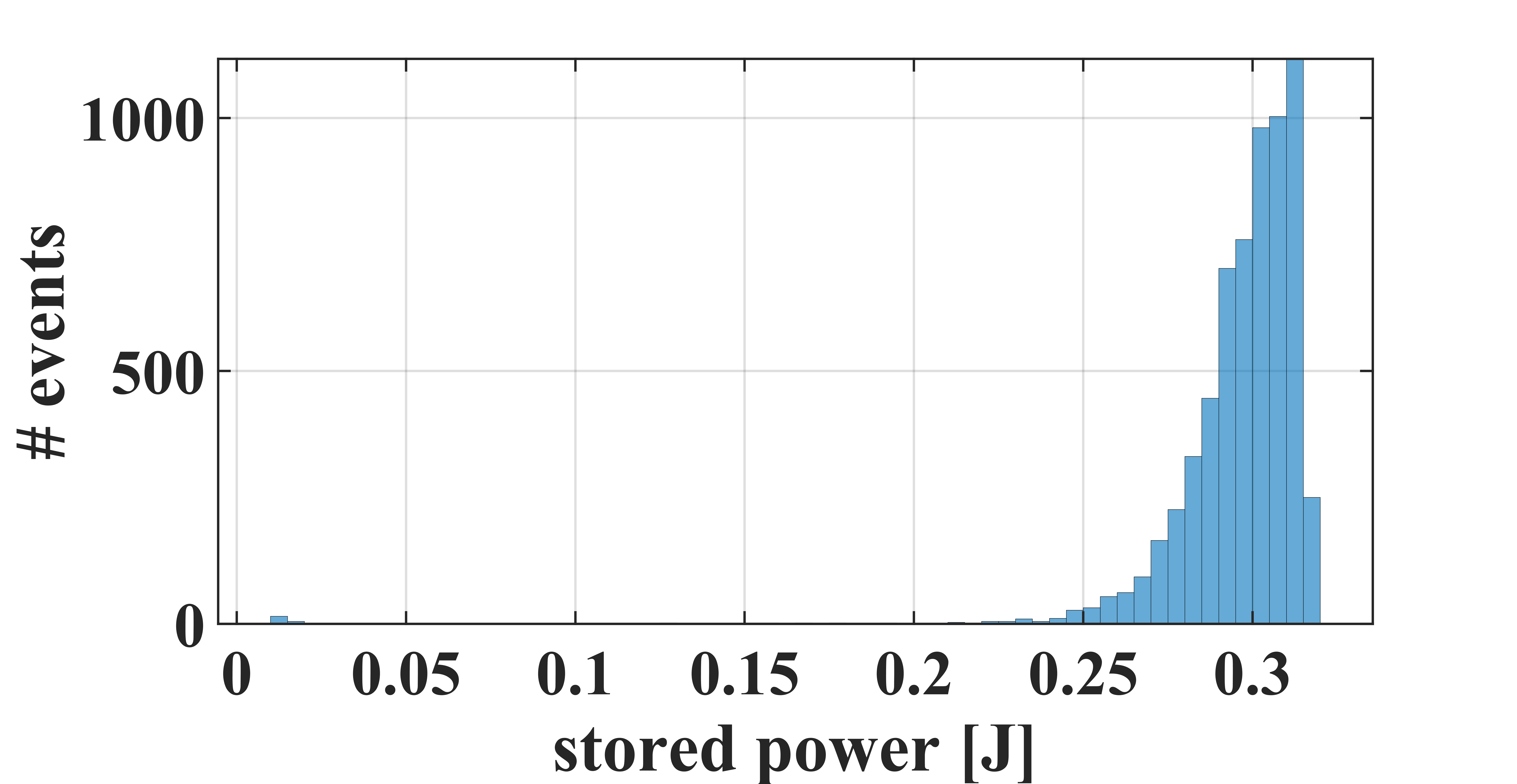}
                    \label{sfig:longterm_histo}}
                \end{minipage}
                \caption{Long-term measurements of energy storage during repeated injection of millijoule-level burst trains. \subref{sfig:longterm_injected_burst} Normalized energy profile of the injected burst. \subref{sfig:longterm_stored_burst} Normalized transmitted burst profile measured at the cavity output. \subref{sfig:longterm_stored} Burst-to-burst energy stability measured over a 20-minute acquisition. \subref{sfig:longterm_histo} Histogram of the stored energy distribution. Energy profiles are smoothed over 50 pulses to reduce pulse-to-pulse photodiode response noise.}
                \label{fig:longterm}
            \end{figure}

        A burst train of 5~$\mu$s duration, corresponding to 4400 laser pulses at 880~MHz, was injected at a repetition rate of 1~kHz. The laser pulse duration was adjusted by means of the compressor stage inside the TANGOR amplifier to approximately 3~ps FWHM (full width at half maximum). Figure~\ref{sfig:longterm_injected_burst} shows the intensity profile of the injected burst train. Its energy distribution primarily reflects the dynamic response of the TANGOR amplifier, exhibiting a high-energy leading edge followed by an exponential decay. As a result, the stored power, monitored via the transmission through mirror $P_2$ and shown in Fig.~\ref{sfig:longterm_stored_burst}, represents the injected profile filtered by the cavity’s low-pass filter response as expected from the finite cavity filling time.

        The injected laser pulse train energy was gradually increased until reaching 2~mJ (integrated over the full burst), corresponding to an average pulse energy of approximately 450~nJ (about 2~W at 1~kHz repetition rate). A long-term storage run was then conducted over 20 minutes. The energy stability is presented in Fig.~\ref{sfig:longterm_stored}, along with its histogram in Fig.~\ref{sfig:longterm_histo}.

        The run was eventually terminated due to instability in the feedback system. At that time, the influence of noise introduced by the TANGOR line on the Lockline signal, due to the polarization extinction ratio between the both lines (see Sec.~\ref{sec:experimental_setup}.\ref{ssec:Experimental_considerations}), had not yet been fully addressed. It might also explain the observed fluctuation on Fig.~\ref{sfig:longterm_stored}. Importantly, no significant thermal effects were noted during the run, apart from a slight drift in average stored power, likely due to environmental changes (room temperature, alignment, etc.). If thermal loading effects, modal instabilities, or changes in the cavity eigenmode had been significant, a variation of the stored power would be expected, as shown in Refs.~\cite{Amoudry:20,710kW}. However, we did not observe any evidence of these types of thermal effects.

        Overall, these results are promising for accelerator-driven Compton X-ray source applications. The stored power showed a stability of approximately 10\%~FWHM (7.6\% root mean square jitter), including the drift, over the 20-minute period. The total stored energy was estimated at approximately 300~mJ (i.e., about 68~$\mu$J per pulse on average). This estimation was based on the gain determined in continuous mode operation and an estimated 60\% coupling via the reflected signal photodiode ($PD_\text{refl}$), see discussion in section~\ref{sec:CEP_experiment}.\ref{ssec:CEP_theory}. The coupling parameter $\Delta_\text{res}$ was measured using the reflected signals from the burst line. The spatial coupling factor $\mathcal{C}$ was then extracted by accounting for the impedance-matching contribution determined from the $T_1$ and $\mathcal{F}$ values and assuming no CEO effect. The uncertainty on the effective gain, and consequently on the estimated stored power, is approximately 15\%. One can note that the telescope used for mode matching was not fully optimized for these experiments. Hence, the laser energy available for Compton scattering is comparable to the maximum energy targeted in the STAR project~\cite{STAR_2016}, but can be achieved without relying on an ultra-high-power laser system and without approaching the onset of nonlinear Compton scattering effects.

    \subsection{Transition to a high-finesse cavity}
        During the low-finesse tests, the coupling mirror was damaged at an input energy of approximately 10~mJ integrated over the full burst (approximately 2~$\mu$J per pulse on average). This mirror had been in use since the first burst-mode experiments conducted in 2016. Thus, we suspect that contamination accumulated on the surface may have reduced the laser-induced damage threshold. We therefore decided to replace all the cavity mirrors with new mirrors originating from the same two manufacturing batches as the previous set: one batch for the end mirrors and one for the coupling mirrors.

        The finesse of the new cavity thereby obtained was measured at $7300 \pm 130$, confirming degradation (likely due to dust contamination) of the older optics. The higher finesse increased the sensitivity of the system to perturbations and required tighter feedback control. In particular, we observed a strong sensitivity of the coupling parameter to the CEO shift, as described in the next section (Sec.~\ref{sec:CEP_experiment}).

\section{CEO dependence of burst power stacking}\label{sec:CEP_experiment}
    This section investigates the influence of carrier-envelope offset (CEO) effects on the energy stacking process inside the optical cavity~\cite{udem_comb}. The cavity filling dynamics are governed not only by the cavity round-trip length, controlled through the locking system, but also by the pulse-to-pulse CEO phase shift of the injected laser field. A theoretical framework is first developed to describe the resulting burst filling pattern and its dependence on the CEO parameters. The model is then confronted with experimental measurements, highlighting the importance of accurate CEO control for achieving optimal power enhancement and preserving the temporal shape of the stored burst.

    \subsection{Model of cavity filling} \label{ssec:CEP_theory}
        The pulse stacking process in a burst-mode optical cavity is governed by two independent parameters: the period detuning between cavity round-trip and laser pulse repetition periods $\delta\tau = \tau_{\text{cav}} - \tau_{\text{rep}}$ and the CEO shift detuning between the CEO shift from pulse to pulse~\cite{udem_comb} and the extra phase shift added by the cavity (accounting for mirror's coating phase shift and Gouy phase shift) $\Delta\Phi_{\mathrm{ce}} = \Phi_{\text{ce}} - \Phi_{\text{cav}}$. The complete derivation of the cavity filling model is given in Appendix~\ref{app:model}.

        The normalized stored energy after $N$ injected pulses is defined as
        \begin{equation}
        \mathcal{R}_{c,N} = \frac{E_{c,N}}{E_{c,\infty}\left(\delta\tau=0,\Delta\Phi_{\mathrm{ce}}=0\right)},
        \end{equation}
        where $E_{c,N}$ is the energy stored inside the cavity after $N$ pulses, and $E_{c,\infty}\left(\delta\tau=0,\Delta\Phi_{\mathrm{ce}}=0\right)$ denotes the maximum steady-state energy that can be stored in the cavity under perfect resonance conditions, i.e. in the absence of both cavity round-trip detuning and CEO phase shift.

        The resonance condition is defined as the configuration for which $\mathcal{R}_{c,\infty}$ reaches its maximum value. This occurs when
        \begin{equation}
        \Delta\Phi_{\mathrm{ce}} + 2\pi f \delta\tau = 2k\pi,\quad k\in\mathbb{Z},
        \end{equation}
        particularly for optical frequencies $f$ around the central frequency of the laser spectrum. Under this resonance condition, the coupling parameter can be written as
        \begin{align}
            \Delta_\text{res} = & \mathcal{C}\left(\left(\frac{R_1 - \left(R_1 + T_1\right) \rho^2}{R_1^2}\frac{T_1}{\left(1-\rho \right)^2}\right) \right.\nonumber \\
            &\left.\frac{\int_{-\infty}^{+\infty}\frac{S(f)}{1 + \frac{4\rho}{(1 - \rho)^2}\sin^2\left(\pi\left(f-\nu_0\right)\frac{f_\text{ce}\tau_\text{rep}}{\nu_0}\right)}\diff f}{\int_{-\infty}^{+\infty}S(f) \diff f}-\frac{T_1}{R_1}\right) \nonumber \\
            = & \left(1-\frac{P_\text{c,ref}}{P_\text{in,ref}}\right),\label{eq:CEP_experiment}
        \end{align}
         where $\mathcal{C}$ accounts for spatial (alignment and beam mode) and polarization coupling imperfections, $S(f)$ is the laser optical spectrum, $\rho = \prod_{i=1}^m r_i$ the product of the mirror reflectivities (accounting for cavity losses), $T_1$ and $R_1$ respectively the intensity transmission and reflection coefficient of the input mirror, $f_{\text{ce}}$ is the CEO frequency shift defined via $\left(\Delta\Phi_{\text{ce}} = 2\pi f_{\text{ce}} \tau_{\text{rep}} \mod{2\pi}\right)$, and $P_\text{c,ref}$ and $P_{\text{in,ref}}$ denote respectively the reflected powers at and far from resonance. The full derivation of Eq.~(\ref{eq:CEP_experiment}) is given in Appendix~\ref{app:model}.

    \subsection{Experimental CEO scan}
        To optimize the stored energy in the cavity, both $\Delta\Phi_{\text{ce}}$ and the spatial coupling factor $\mathcal{C}$ must be adjusted. However, Eq.~(\ref{eq:CEP_experiment}) shows that these two quantities are coupled through the experimentally accessible parameter $\Delta_\text{res}$.

        To disentangle their respective contributions, we performed a controlled scan of the CEO frequency shift $f_{\mathrm{ce}}$. This was achieved by adjusting the optical frequency offset introduced in the Lockline amplification chain through $\mathrm{AOM}_4$ (see Fig.~\ref{fig:setup_minicav}), using both frequency tuning of the AOM driver and two AOMs operating at different nominal frequencies.

        For each value of $f_{\mathrm{ce}}$, the laser repetition period $\tau_{\text{rep}}$ was swept across the cavity resonance. The reflected signal was recorded using a photodiode placed at the reflection of the input mirror.
        From each resonance scan, we extracted the reflected power at resonance, $P_{c,\mathrm{refl}}$, corresponding to the minimum of the reflection curve, and the reflected power far from resonance, $P_{in,\mathrm{refl}}$, corresponding to its maximum.

        From these extrema, we computed the normalized reflection ratio $\frac{P_\text{c,ref}}{P_\text{in,ref}}$, which serves as an experimental observable. The ratio was then fitted using Eq.~(\ref{eq:CEP_experiment}) to retrieve the parameters of interest. In the fitting procedure, the only free parameters were the spatial coupling coefficient $\mathcal{C}$ and the global offset of the CEO frequency $f_{\text{ce}}$. All other parameters were independently measured or derived from prior measurements. Specifically, the optical spectrum $S(f)$ was experimentally acquired and showed a nearly Gaussian profile centered at $\lambda_0 = 1033$~nm with a standard deviation of $\sigma_\lambda = 2.5$~nm. We have checked that the result is not modified significantly when using the true measured spectrum rather than Gaussian. For the simulation we took the input mirror transmission $T_1 = 590 \mathrm{ppm}$, the laser repetition period $\tau_{\text{rep}} = 1/880$~MHz, and the cavity finesse $\mathcal{F} = 7300$, from which the reflectivities $R_1 = 1-T_1$ and $\rho = 1-\frac{\pi}{\mathcal{F}}$ were calculated.

\begin{figure}[htb!]
            \centering
            \includegraphics[width=\linewidth]{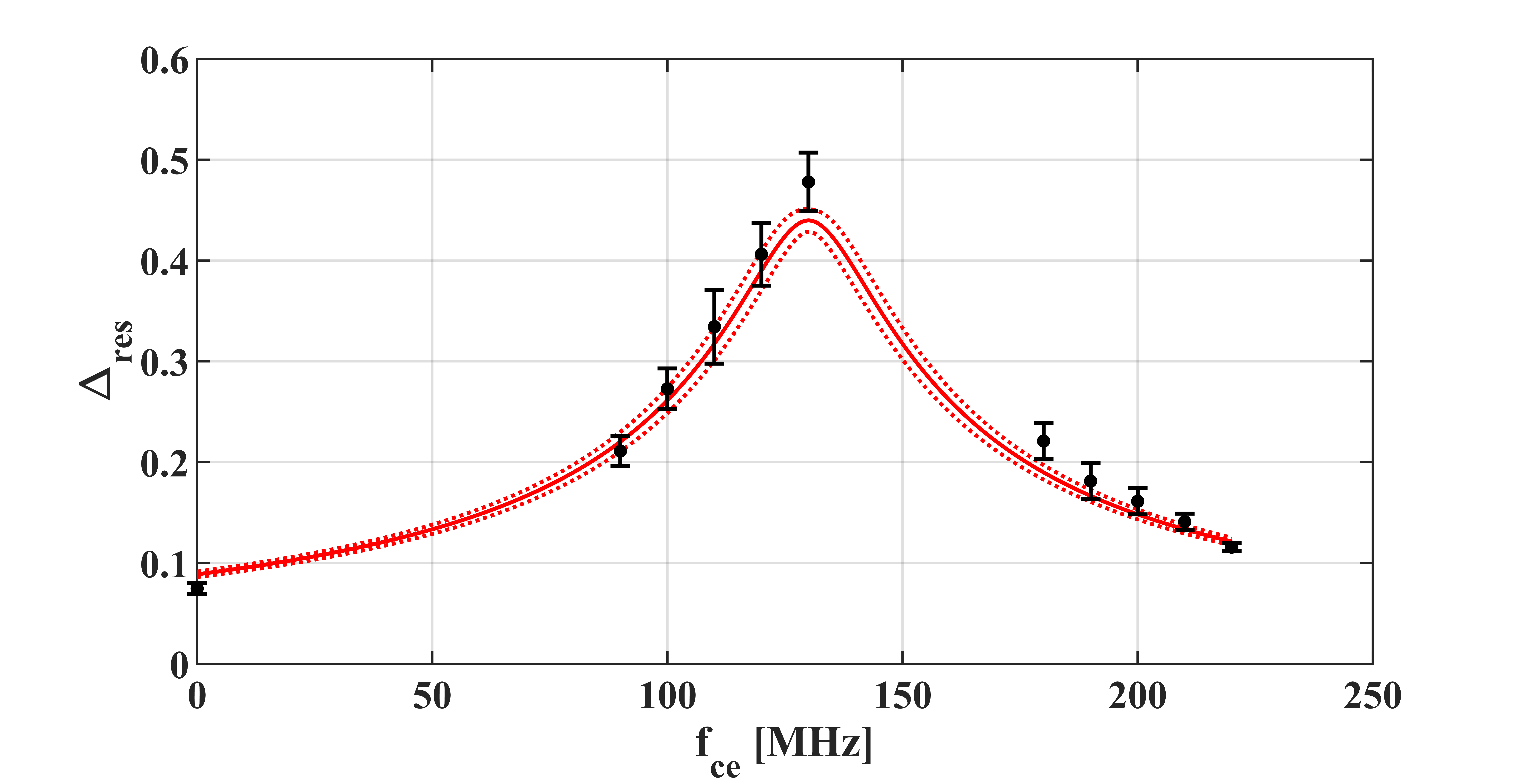}
            \caption{Results of the cavity $f_\text{ce}$ scan. Black dots represent $\left(1-\frac{P_\text{c,ref}}{P_\text{in,ref}}\right)$, with error bars estimated from the fluctuation of background and repeatability over 10 measurements. The solid red line shows the fit of the data using the cavity coupling model described in Eq.~\ref{eq:CEP_experiment}, and dashed red lines represent the 63\% confidence bounds.}
            \label{fig:CEP_scan_results}
        \end{figure}

        Figure~\ref{fig:CEP_scan_results} presents the experimental data points together with the best-fit theoretical curve derived from equation~\ref{eq:CEP_experiment}. The model accurately reproduces the measurements across the entire range of $f_{\text{ce}}$ values. The confidence interval associated with the fit encompasses nearly all the measured data points, indicating a good agreement between theory and experiment. The extracted parameters from the fit are $\mathcal{C} = 0.51 \pm 0.03$ for the spatial coupling and an optimal CEO frequency shift of $f_{\text{ce}} = 130 \pm 5 \mathrm{MHz}$.

        This analysis demonstrates that the model can reliably predict the cavity injection efficiency and enables a robust estimation of the spatial mode coupling as well as the value for a $\Delta\Phi_{\text{ce}}$ required for optimal energy build-up in the cavity.

        The measurements presented above provide access to the CEO frequency shift associated with the locking beam and allow the optimization of the cavity coupling. However, the CEO frequency shift experienced by the burst beam may differ from that of the Lockline because the two beams propagate through independent amplification chains. As a consequence, even when the cavity is maintained at resonance by the locking system, a residual CEO frequency mismatch can remain between the Lockline and the burst beam. In the following section, we investigate the impact of this mismatch on the burst stacking dynamics.

    \subsection{Sensitivity to \texorpdfstring{$f_{\text{ce}}$}{f₍ce₎} and \texorpdfstring{$\Delta\Phi_\text{ce}$}{ΔΦ₍ce₎}}
        Two distinct CEO-related quantities must therefore be considered. The first one is the absolute CEO frequency shift, $f_\text{ce}$, which governs the spectral overlap between the laser comb and the cavity modes. The second one is the CEO frequency mismatch, $\Delta f_\text{ce}$, between the Lockline and the burst amplification chain. While the locking system continuously compensates the effect of $f_\text{ce}$ to maintain cavity resonance, it cannot compensate for $\Delta f_\text{ce}$ because the latter originates from differences between the two optical paths. Consequently, a pulse-to-pulse phase error accumulates during the burst injection process and may strongly affect the achievable stored energy.

        First, in order to illustrate the influence of the absolute CEO phase $\Delta\Phi_\text{ce} = 2\pi f_\text{ce}$ on the intracavity power buildup, we numerically simulate the cavity response for various values of $\Delta\Phi_\text{ce}$, assuming that the cavity is kept at resonance, i.e., such that the phase term $(\Delta\Phi_{\text{ce}} + 2\pi f \delta\tau) = 0$. Two different cavity finesses are considered: a low finesse ($\mathcal{F} = 2000$) and a high finesse ($\mathcal{F} = 7300$).

\begin{figure}[htb!]
            \centering
            \begin{minipage}{\columnwidth}
                \centering
                \subfigure[]{
                \includegraphics[width=\linewidth]{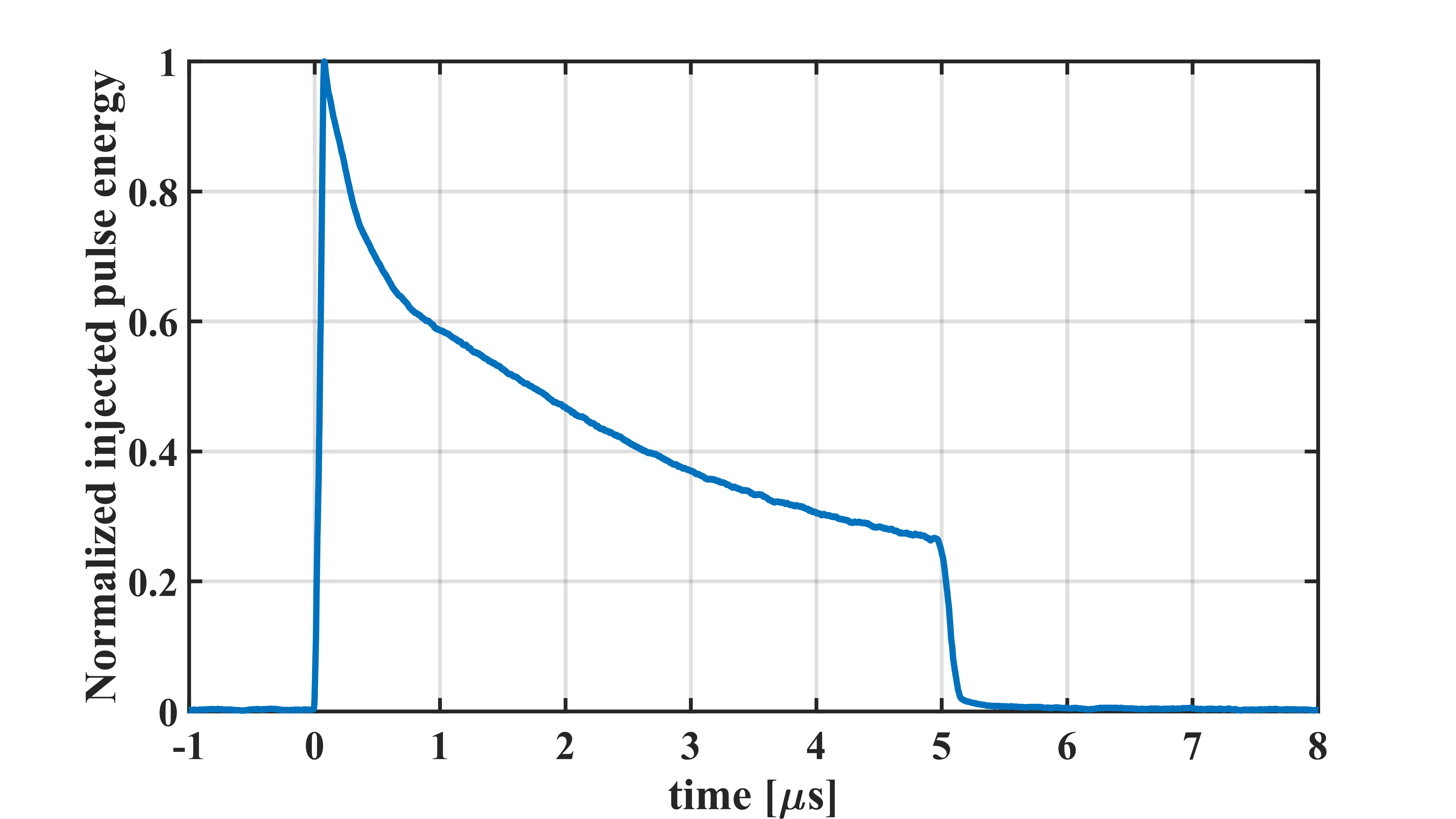}
                \label{sfig:burst_injected_F7300}}
            \end{minipage}\\
            \begin{minipage}{\columnwidth}
                \centering
                \subfigure[]{
                \includegraphics[width=\linewidth]{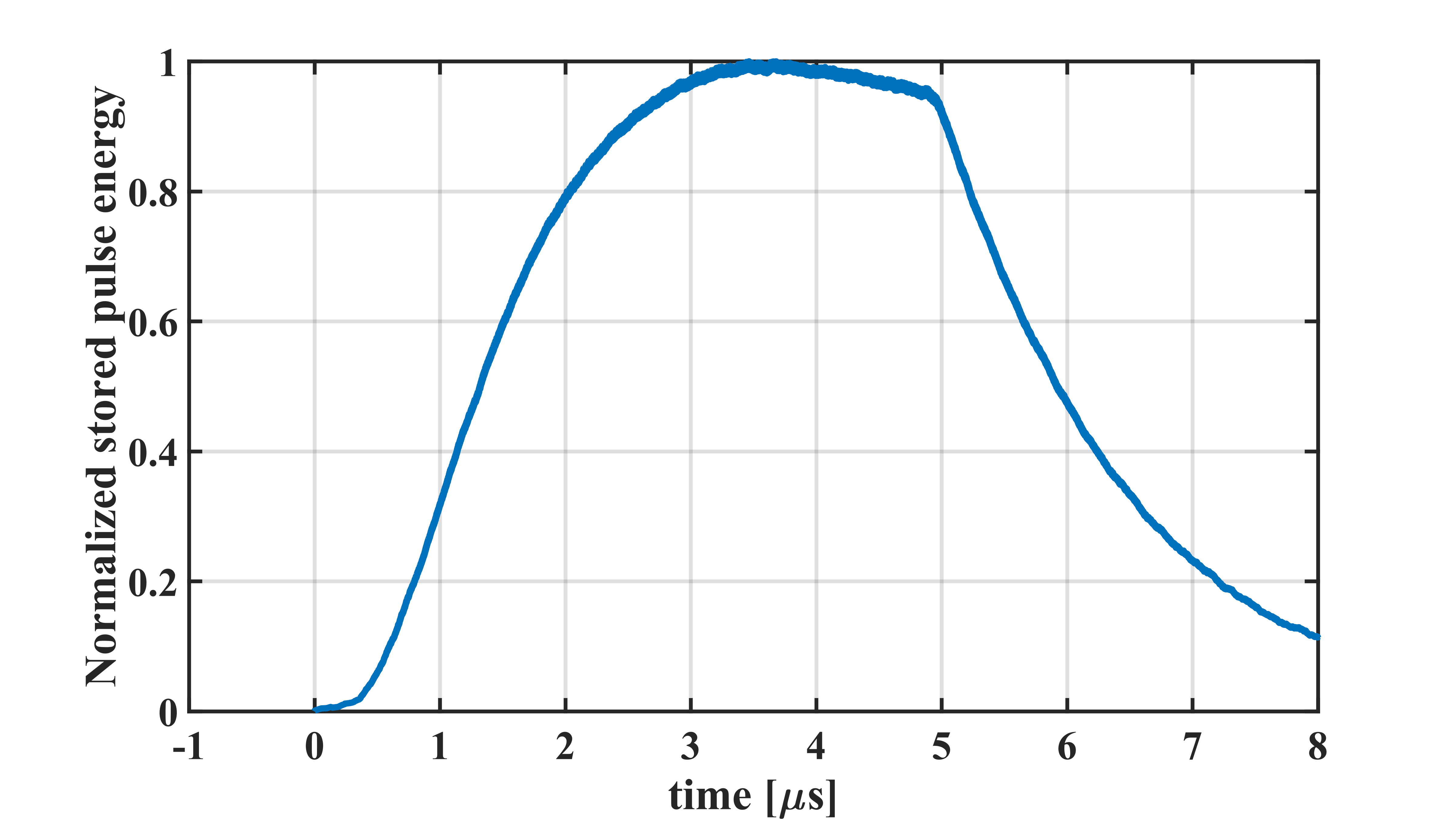}
                \label{sfig:burst_stored_F7300}}
            \end{minipage}
            \caption{Measured burst energy profiles for the new cavity ($\mathcal{F} = 7300$). \subref{sfig:burst_injected_F7300} Normalized energy profile of a typical injected burst; \subref{sfig:burst_stored_F7300} Measured normalized transmitted energy profile through the cavity. Both profiles were obtained using the same numerical processing as in Fig.~\ref{fig:longterm}.}
            \label{fig:burst_F7300}
        \end{figure}

        We modify the expression of the intracavity energy after the injection of $N$ pulses to account for the energy of each pulse in the burst. The energy stored in the cavity becomes:
        \begin{multline}
            E_{c,N} = T_1 \int_{-\infty}^{+\infty} \left| \sum_{k=0}^{N-1} \left[ \rho \exp{\left(-\im \left(2\pi \nu_0 \delta\tau + \Delta\Phi_{\text{ce}}\right)\right)} \right]^k \cdot \right. \\
            a_k(t - k\delta\tau) \biggr|^2 \diff t,
        \end{multline}
        where $a_k(t)$ denotes the temporal envelope of the $k$-th pulse in the burst, modeled as a Gaussian pulse of amplitude $A_k$, defined as:
        \begin{equation}
            a_k(t) = A_k \left( \frac{2\pi \sigma_f}{\sqrt{\pi/2}} \right)^{1/2} \exp\left( -4\pi^2 \sigma_f^2 t^2 \right),
        \end{equation}
        with $\sigma_f = \frac{\sigma_\lambda c}{\lambda_0^2}$ the spectral standard deviation of the optical pulse (assuming a Gaussian optical spectrum). The values $A_k$ are extracted from the experimentally measured burst profile normalized to a unit maximum, shown in Fig.~\ref{sfig:burst_injected_F7300}.

\begin{figure}[htb!]
            \centering
            \begin{minipage}{\columnwidth}
                \centering
                \subfigure[]{
                \includegraphics[width=\linewidth]{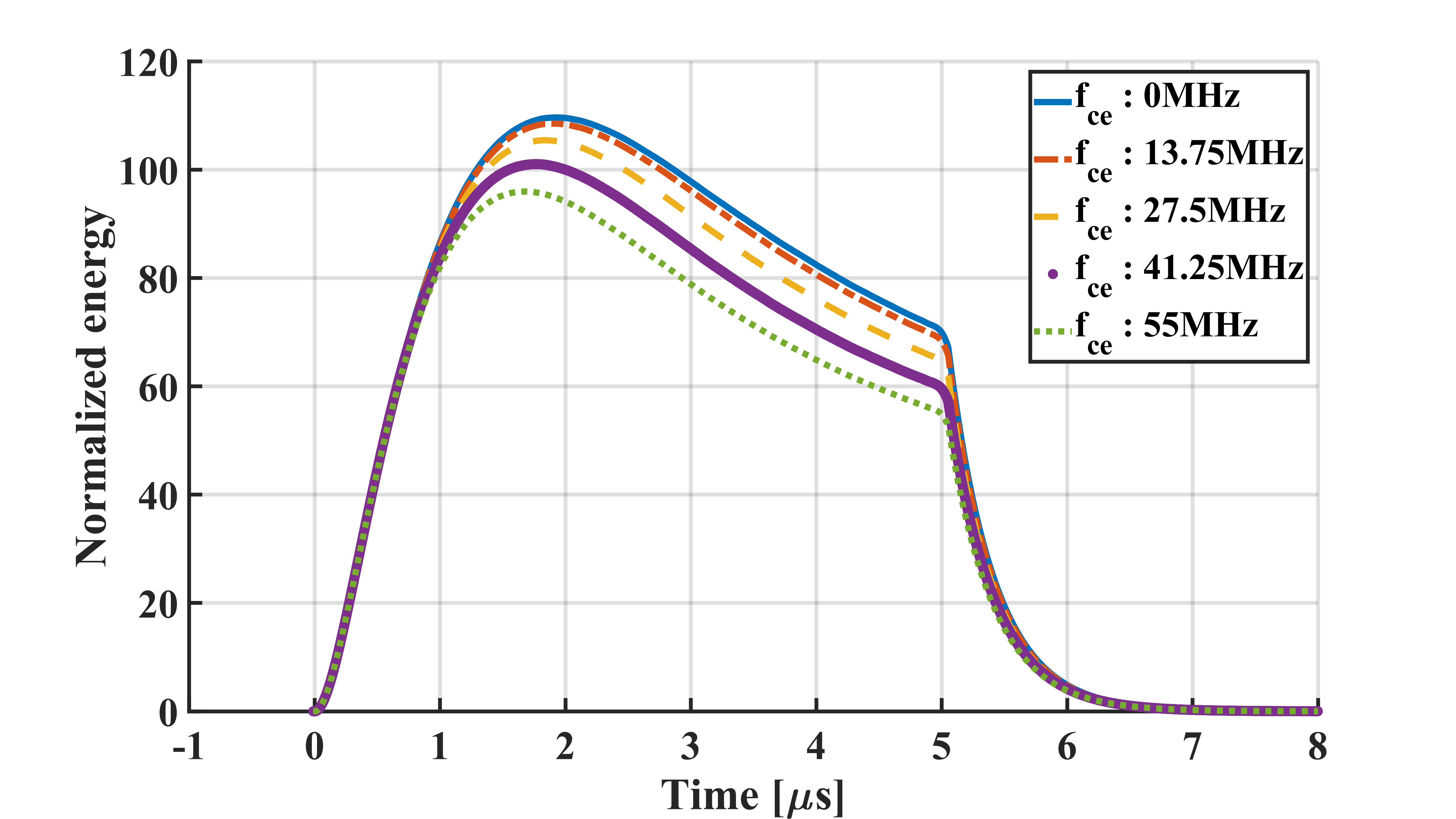}
                \label{sfig:CEP_effect_F2000}}
            \end{minipage}\\
            \begin{minipage}{\columnwidth}
                \centering
                \subfigure[]{
                \includegraphics[width=\linewidth]{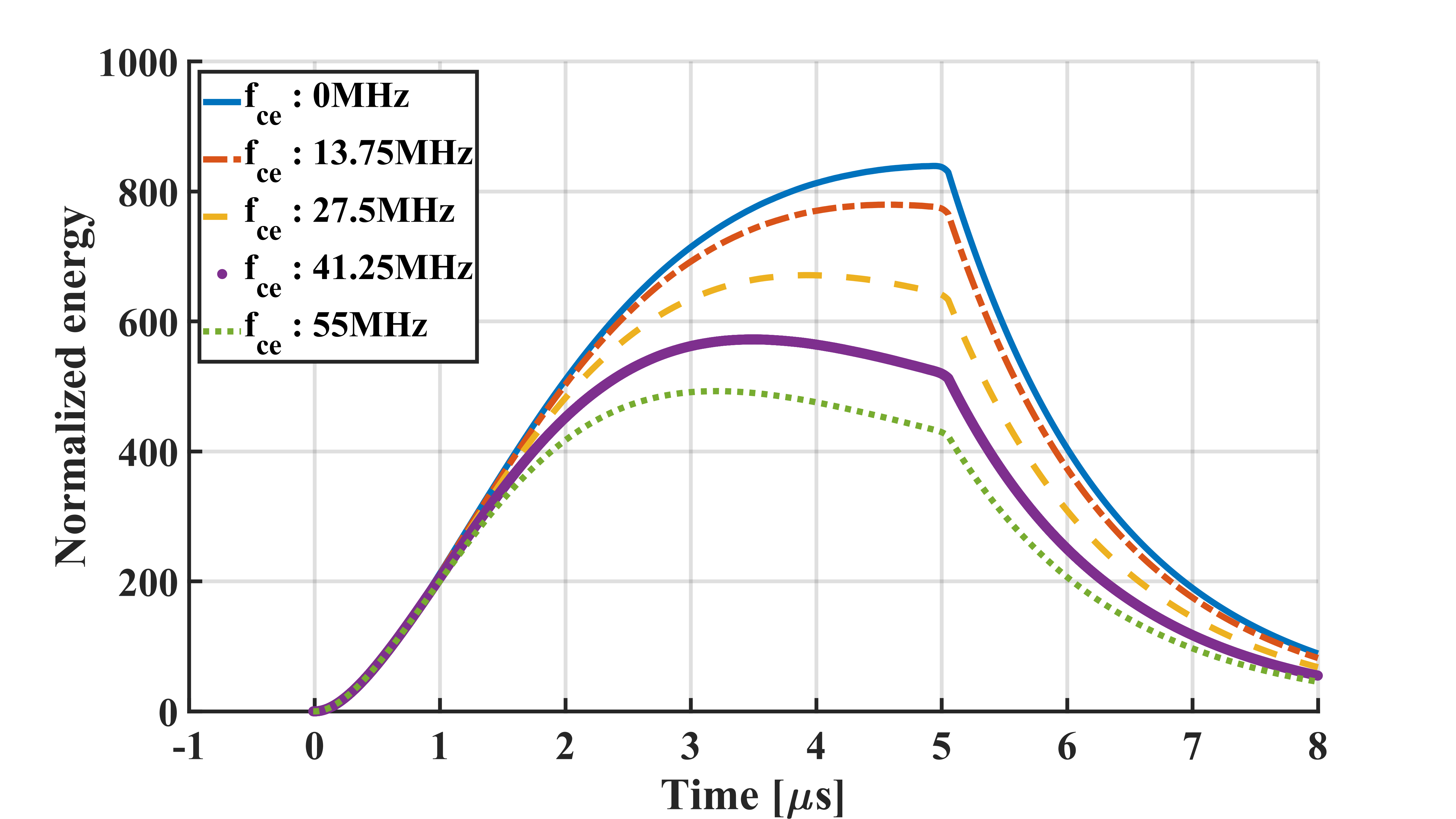}
                \label{sfig:CEP_effect_F7300}}
            \end{minipage}
            \caption{Effect of $f_\text{ce}$ on the energy profile inside the cavity of a burst of laser pulses for two finesse values: \subref{sfig:CEP_effect_F2000}  $\mathcal{F} = 2000$, \subref{sfig:CEP_effect_F7300} $\mathcal{F} = 7300$. Simulations were performed with the cavity set at resonance.}
            \label{fig:CEP_effect}
        \end{figure}

        The results of the simulations are presented in Fig.~\ref{fig:CEP_effect}. As expected, the influence of $\Delta\Phi_\text{ce}$ is significantly more pronounced in the high finesse configuration. In particular, the simulation for $\mathcal{F} = 7300$ exhibits a qualitative agreement with the experimentally measured profile of Fig.~\ref{sfig:burst_stored_F7300}, provided a detuning of $\Delta\Phi_{\text{ce}} \gtrsim 41$~MHz is assumed. In contrast, the simulations for the low finesse case (Fig.~\ref{sfig:CEP_effect_F2000}) fail to reproduce the features observed in the experimental profile of Fig.~\ref{sfig:longterm_stored_burst}.

        This discrepancy suggests that additional effects, not captured in the above model, play a role. Notably, the simulation assumes a common CEO frequency $f_{\text{ce}}$ for both the Lockline and the burst line. In practice, there is no reason to expect these two $f_{\text{ce}}$ to be matched: the AOMs used to introduce a frequency shift in the TANGOR chain are only partially compensated by an AOM in the Lockline. Moreover, each amplification chain has its own dynamics, potentially leading to additional uncompensated CEO shifts.

        Additionally, the locking system acting on the Lockline, to maintain the resonance condition, introduces a feedback mechanism that can partially compensate its own CEO frequency. As a result, the residual mismatch $\Delta f_{\text{ce}}$ between the two CEO frequencies persists. This mismatch leads to a temporal accumulation of phase shift between the electric fields of the Lockline and the burst line. To account for this effect, we update the previous model by including the term $2\pi \Delta f_\text{ce} \tau_\text{rep}$ in the total CEO phase shift, yielding:
        \begin{equation}
            \Delta\Phi_\text{ce} \rightarrow \Delta\Phi_\text{ce} + 2\pi \Delta f_\text{ce} \tau_\text{rep}.
        \end{equation}

\begin{figure}[htb!]
        \centering
        \begin{minipage}{\columnwidth}
            \centering
            \subfigure[]{
            \includegraphics[width=\linewidth]{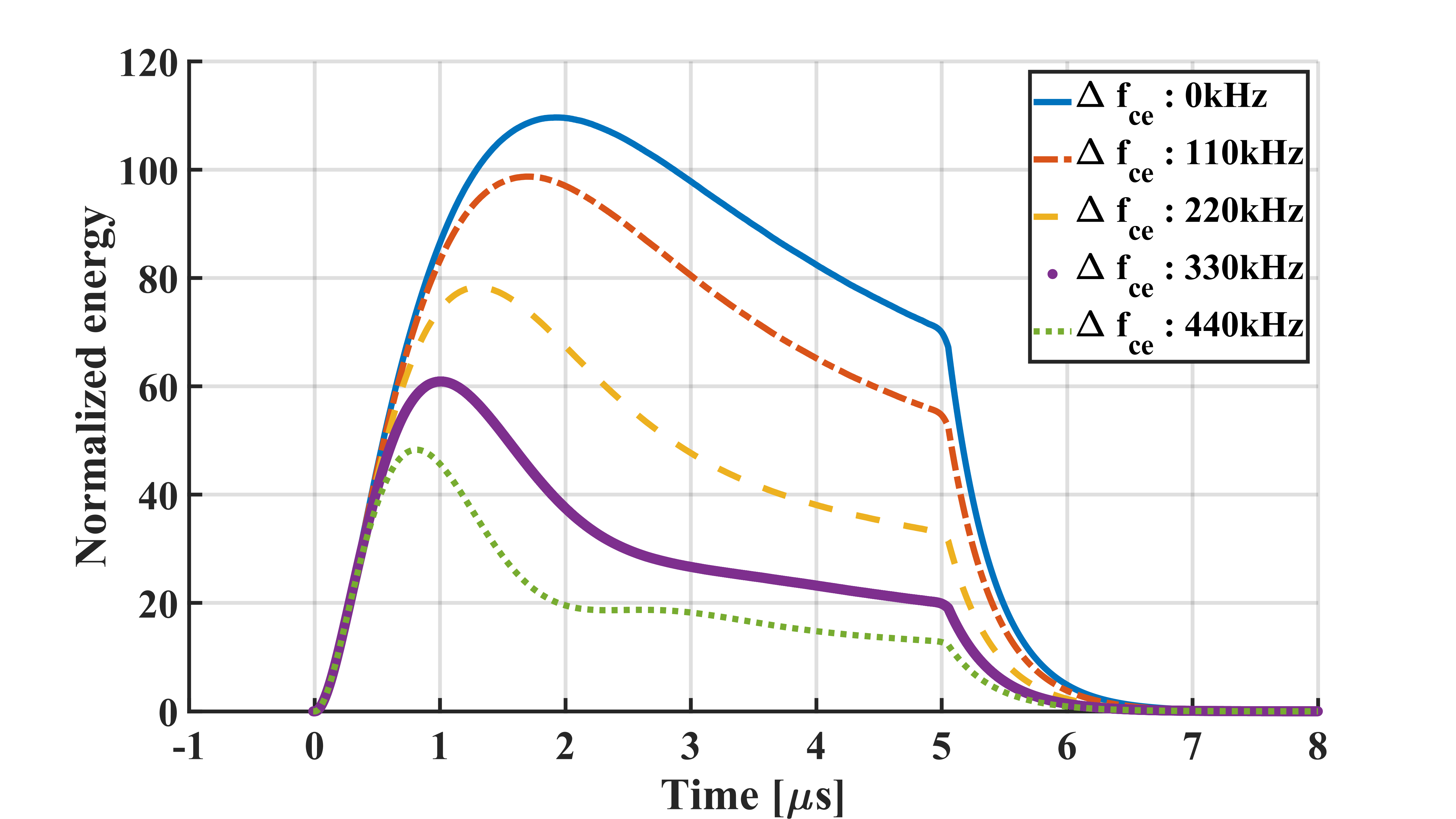}
            \label{sfig:CEPSlip_effect_F2000}}
        \end{minipage}\\
        \begin{minipage}{\columnwidth}
            \centering
            \subfigure[]{
            \includegraphics[width=\linewidth]{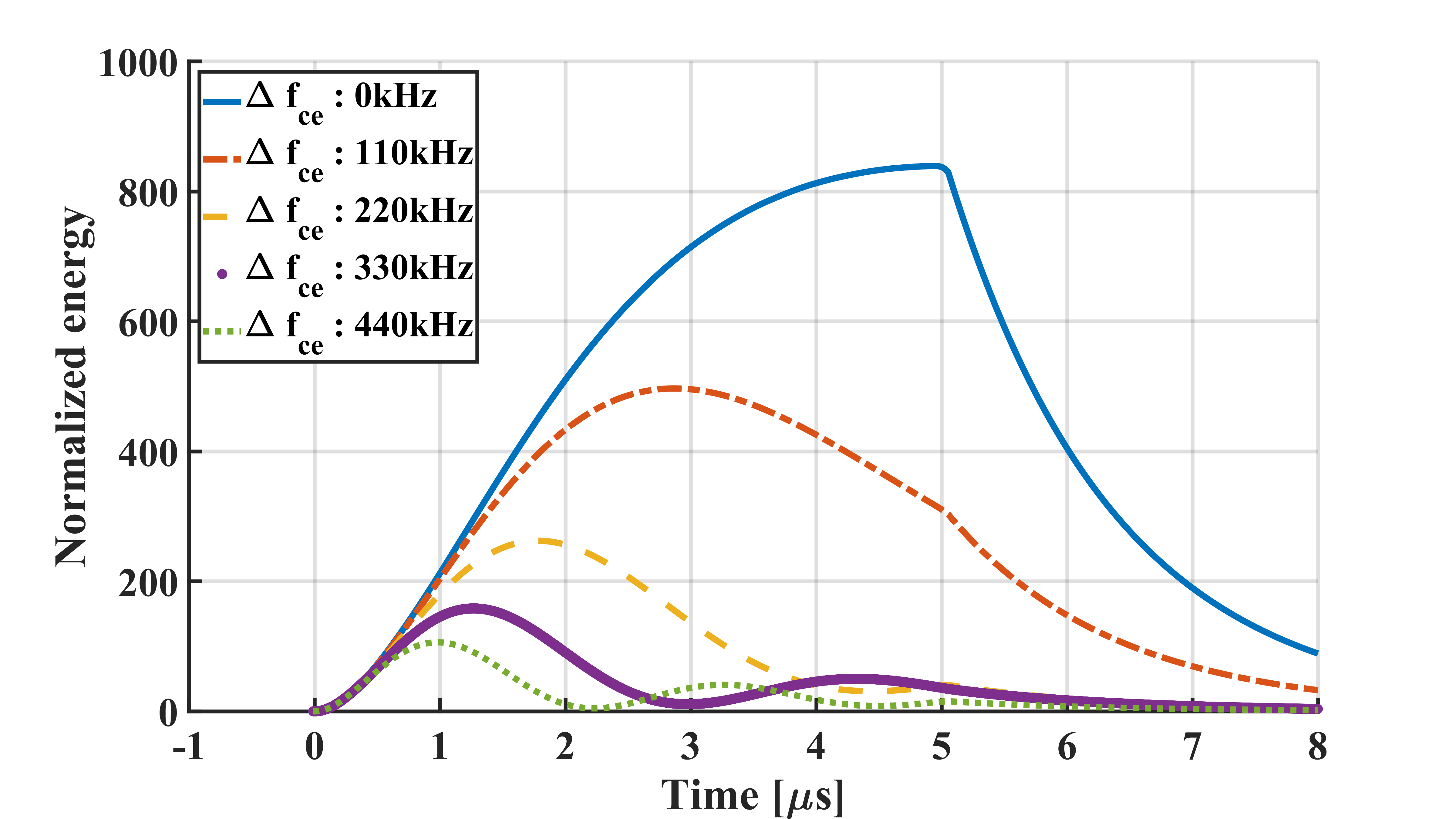}
            \label{sfig:CEPSlip_effect_F7300}}
        \end{minipage}
        \caption{Effect of a detuning $\Delta f_\text{ce}$ on the energy profile inside the cavity of a burst of laser pulses for two finesse values: \subref{sfig:CEP_effect_F2000}  $\mathcal{F} = 2000$, \subref{sfig:CEP_effect_F7300} $\mathcal{F} = 7300$. Simulations were performed with $f_\text{ce} = 0$ and the cavity set at resonance.}
        \label{fig:CEPSlip_effect}
    \end{figure}

        We perform a second series of simulations by setting $\Delta\Phi_\text{ce} = 0$ and introducing only a nonzero $\Delta f_\text{ce}$. The results are shown in Fig.~\ref{fig:CEPSlip_effect}. The simulations clearly reveal that variations in $\Delta f_\text{ce}$ have a far more pronounced effect on the maximum stored energy than equivalent variations in $\Delta\Phi_\text{ce}$, with differences spanning several orders of magnitude. Furthermore, the shape of the stored energy profile is highly sensitive to this mismatch. In Fig.~\ref{sfig:CEPSlip_effect_F2000} curves for a detuning of $\Delta f_\text{ce} \gtrsim 220$~kHz show qualitative feature of the experimental profile of Fig.~\ref{sfig:longterm_stored_burst}. Better agreement can be obtained by changing  at the same time both $\Delta f_\text{ce}$ and $\Delta\Phi_\text{ce}$. Note that no dedicated optimization was made during this first experiment.

        No quantitative fit of the energy profile was performed because the available data were not acquired in a dedicated measurement campaign for this purpose. A dedicated experiment would be required to quantify the contribution of the acquisition system, including the photodiode and oscilloscope responses, as well as residual power leakage, to the measured profile. This is kept out of the scope of this paper.

\section{Conclusion}
    In this work, we have demonstrated the feasibility of efficiently storing high-energy laser bursts in a passive optical cavity operated in burst mode. Using a cavity with a repetition frequency of 880~MHz and composed entirely of commercial mirrors, we were able to store more than 300~mJ over periods exceeding 20 minutes, without applying any in-depth optimization.

    The model of pulse stacking developed in this paper shows that the CEO shift and the coupling parameter are entangled. This indicates that measuring the coupling parameter requires a careful characterization of the cavity to avoid risk of misestimation. The proposed method based on a scan of the CEO shift allows the determination of the coupling efficiency and the CEO shift of the cavity at the same time. A limitation of the model is that we have not implemented dynamic effect on the CEO shift that might arise in the amplification chain. The modeling of this effect and its experimental evidence remain out of the scope of this paper.

    A key outcome of this work is the identification of two distinct CEO-related quantities governing the burst stacking process. The first one, $f_\mathrm{ce}$, determines the spectral matching between the laser frequency comb and the cavity resonances. The second one, $\Delta f_\mathrm{ce}$, corresponds to the CEO frequency mismatch between the Lockline and the burst amplification chain. While the locking system can compensate for variations of $f_\mathrm{ce}$ by maintaining cavity resonance, it does not suppress $\Delta f_\mathrm{ce}$. Our simulations show that both quantities play a central role in the energy stacking process and that their influence increases significantly as the cavity finesse increases. Importantly, the impact of $\Delta f_\mathrm{ce}$ on the achievable stored energy and on the burst temporal profile is found to be substantially larger than that of $f_\mathrm{ce}$, making it a critical parameter for future high-finesse burst enhancement cavities.

    For a cavity with a finesse on the order of several thousand, maintaining energy losses below a few percent requires tuning $f_\text{ce}$ to within a few tens of megahertz of the repetition frequency (880~MHz). The constraint on $\Delta f_\text{ce}$ is even more stringent: to keep losses similarly low, it must be stabilized within a few tens of kilohertz (about the linewidth of the cavity).

    These findings pave the way for further enhancement of the stored energy in burst-mode optical cavities. Precise control and tuning of CEO parameters will be essential in future experiments aiming at joule-level energy storage. Importantly, no limiting effects such as laser-induced damage or thermal degradation were observed during our tests, suggesting that achieving 1~J of stored energy is within reach. Future work will focus on active control of CEO-related parameters and on scaling the stored burst energy toward the joule level required for next-generation burst-mode inverse Compton sources.

\appendix
\section{Derivation of the cavity filling model} \label{app:model}
    We consider the electric field of a pulsed laser expressed as:
    \begin{align*}
        e(t) = & \sum_{q=-\infty}^{+\infty} e_q(t) \\
             = & \sum_{q=-\infty}^{+\infty} a(t - q\tau_{\text{rep}}) \exp\left[\im \left(2\pi\nu_0(t - q\tau_{\text{rep}}) + q\Phi_{\text{ce}}\right)\right],
    \end{align*}
    where $e_q(t)$ is the electric field of the $q$-th pulse, $\nu_0$ is the optical carrier frequency, $a(t)$ is the temporal envelope of a single pulse, $\tau_{\text{rep}}$ is the pulse repetition period, and $\Phi_{\text{ce}}$ is the CEO shift from pulse to pulse~\cite{udem_comb}.

    The electric field stored in a Fabry–Perot cavity after injecting $N$ consecutive pulses is:
    \begin{equation*}
        e_{c,N}(t) = \im t_1 \sum_{k=0}^{N-1} \left(\rho \exp\left[\im\Phi_{\text{cav}}\right]\right)^k e_{N-1-k}(t - k \tau_{\text{cav}}),
    \end{equation*}
    with $\rho = \prod_{i=1}^m r_i$ the product of the mirror reflectivities (accounting for cavity losses), $\Phi_{\text{cav}}$ the extra phase shift added by the cavity (accounting for mirror's coating phase shift and Gouy phase shift), $t_1$ the transmission of the input mirror, and $\tau_{\text{cav}}$ the cavity round-trip time. We consider for the moment only longitudinal dynamics, and neglect also transverse effects.

    Let $\tilde{X}(f)$ denote the Fourier transform of $X(t)$. Then, in the frequency domain, the stored electric field reads:
    \begin{multline*}
        \tilde{e}_{c,N}(f) = \tilde{a}(f - \nu_0) \exp\left[\im\left(N-1\right)\left(\Delta\Phi_{\text{ce}} - 2\pi f \tau_{\text{rep}}\right)\right] \cdot \\
        \frac{\im t_1 \left[1 - \rho^N \exp\left(-\im N\left(\Delta\Phi_{\text{ce}} + 2\pi f \delta\tau\right)\right)\right]}{1 - \rho \exp\left[-\im\left(\Delta\Phi_{\text{ce}} + 2\pi f \delta\tau\right)\right]},
    \end{multline*}
    where $\delta\tau = \tau_{\text{cav}} - \tau_{\text{rep}}$ is the period detuning and $\Delta\Phi_{\text{ce}} = \Phi_{\text{ce}} - \Phi_{\text{cav}}$ is the CEO shift detuning between laser pulse to pulse and cavity phase shift.

    From Parseval’s theorem, the energy stored after $N$ pulses is:
    \begin{equation*}
        E_{c,N} = \int_{-\infty}^{+\infty} \vert e_{c,N}(t)\vert^2 \diff t = \int_{-\infty}^{+\infty} \vert\tilde{e}_{c,N}(f)\vert^2 \diff f.
    \end{equation*}
    We define the normalized energy parameter:
    \begin{equation*}
        \mathcal{R}_{c,N} = \frac{E_{c,N}}{E_{c,\infty}\left(\delta\tau = 0, \Delta\Phi_{\text{ce}} = 0\right)},
    \end{equation*}
    representing the ratio between stored energy after $N$ pulses and the maximum achievable energy (without period and CEO shift detuning).

    Since $\mathcal{R}_{c,N}$ is not directly measurable and sensitive to input power fluctuations, we focus on the reflected power from the input mirror:
    \begin{equation*}
        E_{r,N} = \int_{-\infty}^{+\infty} \left| r_1 e_{N-1}(t) + \im t_1 \frac{\rho}{r_1} e_{c,N-1}\left(t - \tau_{\text{cav}}\right) \right|^2 dt,
    \end{equation*}
    with $r_1$ the reflectivity of the input mirror, and we define:
    \begin{equation*}
        \mathcal{R}_{r,N} = \frac{E_{r,N}}{E_0},
    \end{equation*}
    with $E_0$ the incident pulse energy. In the absence of resonance, the reflected energy is $R_1 E_0$, where $R_1$ is the intensity reflection coefficient of the input mirror.

    Thus, we define the coupling parameter:
    \begin{align}
        \Delta = & \mathcal{C}\left(1-\frac{\mathcal{R}_{r,\infty}}{R_1}\right) \nonumber \\
        = & \mathcal{C}\left(\left(\frac{R_1 - \left(R_1 + T_1\right) \rho^2}{R_1^2}\frac{T_1}{\left(1-\rho \right)^2}\right)\mathcal{R}_{c,\infty}-\frac{T_1}{R_1}\right) \nonumber \\
        = & \mathcal{C}\left(\left(\frac{R_1 - \left(R_1 + T_1\right) \rho^2}{R_1^2}\frac{T_1}{\left(1-\rho \right)^2}\right) \cdot \right.  \nonumber \\
        & \frac{\int_{-\infty}^{+\infty}\frac{S(f)}{1 + \frac{4\rho}{(1 - \rho)^2}\sin^2\left(\frac{1}{2} (\Delta\Phi_{\text{ce}} + 2\pi f \delta\tau) \right)}\diff f}{\int_{-\infty}^{+\infty}S(f) \diff f}-\frac{T_1}{R_1}\Biggr),\label{eq:Delta_CEP}
    \end{align}
    where $\mathcal{C}$ accounts for all spatial (alignment and beam mode) and polarization coupling imperfections not considered before, and $S(f)$ is the laser optical frequency spectrum. We define the resonance condition as the situation in which the stored energy ratio $\mathcal{R}_{c,\infty}$ is maximized. This occurs when the phase term $(\Delta\Phi_{\text{ce}} + 2\pi f \delta\tau)$ equals an integer multiple of $2\pi$, particularly around the central frequency of the laser spectrum. When this resonance condition is fulfilled we can derive a new expression for equation~\eqref{eq:Delta_CEP}:
    \begin{align}
        \Delta_\text{res} = & \mathcal{C}\left(\left(\frac{R_1 - \left(R_1 + T_1\right) \rho^2}{R_1^2}\frac{T_1}{\left(1-\rho \right)^2}\right) \cdot \right.  \nonumber \\
        & \frac{\int_{-\infty}^{+\infty}\frac{S(f)}{1 + \frac{4\rho}{(1 - \rho)^2}\sin^2\left(\pi\left(f-\nu_0\right)\frac{f_\text{ce}\tau_\text{rep}}{\nu_0}\right)}\diff f}{\int_{-\infty}^{+\infty}S(f) \diff f}-\frac{T_1}{R_1}\Biggr) \nonumber \\
        = & \left(1-\frac{P_\text{c,ref}}{P_\text{in,ref}}\right),
    \end{align}
    where $f_{\text{ce}}$ is the CEO frequency shift defined via $\left(\Delta\Phi_{\text{ce}} = 2\pi f_{\text{ce}} \tau_{\text{rep}} \mod{2\pi}\right)$, and $P_\text{c,ref}$ and $P_{\text{in,ref}}$ denote the reflected powers at and far from resonance, respectively.

\begin{backmatter}
\bmsection{Funding}
The present work is partially financed by Amplitude Laser company.

\bmsection{Disclosures}
The authors declare no conflicts of interest.

The authors used ChatGPT (from GPT-4.1 to GPT-5.5 versions) solely for grammar refinement and syntax suggestions. The tool was not used to generate scientific claims, references, experimental data, or conclusions.

\bmsection{Data availability} Data underlying the results presented in this paper are not publicly available at this time but may be obtained from the authors upon reasonable request.

\end{backmatter}

\bibliography{biblio}

\end{document}